# Band gap tuning and orbital mediated electron – phonon coupling in HoFe$_{1-x}$Cr$_x$O$_3$ (0 ≤ x ≤ 1)

Ganesh Kotnana[1] and S. Narayana Jammalamadaka[1*]

[1]*Magnetic Materials and Device Physics Laboratory, Department of Physics, Indian Institute of Technology Hyderabad, Hyderabad, India – 502 205.*

*Corresponding author: surya@iith.ac.in

**Abstract:**

We report on the evidenced orbital mediated electron – phonon coupling and band gap tuning in HoFe$_{1-x}$Cr$_x$O$_3$ (0 ≤ x ≤ 1) compounds. From the room temperature Raman scattering, it is apparent that the electron-phonon coupling is sensitive to the presence of both the Fe and Cr at the B-site. Essentially, an A$_g$ like local oxygen breathing mode is activated due to the charge transfer between Fe$^{3+}$ - Cr$^{3+}$ at around 670 cm$^{-1}$, this observation is explained on the basis of Franck-Condon (FC) mechanism. Optical absorption studies infer that there exists a direct band gap in the HoFe$_{1-x}$Cr$_x$O$_3$ (0 ≤ x ≤ 1) compounds. Decrease in band gap until x = 0.5 is ascribed to the broadening of the oxygen $p$ - orbitals as a result of the induced spin disorder due to Fe$^{3+}$ and Cr$^{3+}$ at B-site. In contrast, the increase in band gap above x = 0.5 is explained on the basis of the reduction in the available unoccupied $d$ - orbitals of Fe$^{3+}$ at the conduction band. We believe that above results would be helpful for the development of the optoelectronic devices based on the ortho-ferrites.

**Introduction**:

The band gap engineering may play a significant role in the future spin-photonic and ultra-violet (UV) photonic devices such as laser diodes, solar blind UV photo-detectors light-emitting diodes (LED), and transparent electronic devices. In order to fulfil the above requirement, there is a quest for development of the new materials and engineer the band gap selectively based on the requirement. In search of the new materials, materials based on the rare earth ortho-ferrites $RFeO_3$ (R - rare earth metal and the Fe - transition metal) have appealed continued experimental and theoretical interest as a result of their impressive magnetic and structural changes[1-3]. Technologically these ortho – ferrites have gained much attention in solid oxide fuel cells[4], magneto optic devices[5], gas sensors[6] and for the detection of ozone in monitoring environment[7]. On top of that these compounds also have much potential for the photocatalytic activity[8] and the multiferroic behaviour[9].

Electron – lattice dynamics are very important in the rare earth-transition metal oxides to understand the phenomenon of the colossal magnetoresistance[10]. On the other hand, exploring the correlation between the structural, magnetic and orbital ordering is very much essential to understand the charge transfer mechanism which is base for the electron – phonon coupling[11]. It has been believed that in the cubic perovskites, due to the distortion of the lattice, symmetry of the structure lowers and this leads to the appearance of the Raman-active phonons[12]. The influence of the electronic configuration and orbital ordering on the Raman spectra of these distorted perovskites has been addressed by Allen *et. al.*[13]. Local oxygen breathing mode in the Raman scattering has been observed in the distorted perovskite compound based on the mixed Fe and Cr ions at B – site. Such intriguing phenomenon has been explained on the basis of the Franck-Condon picture following a photon induced transfer of an electron from Fe to adjacent Cr ion[11]. Similar behaviour has been observed on the other compound $La_{1-y}Sr_yMn_{1-x}M_xO_3$ (M=Cr, Co, Cu, Zn, Sc or Ga)[14]. To our knowledge, until now the electron - phonon coupling has been observed only in the Jahn – Teller active compounds. However, it would be very interesting if we can observe the electron - phonon coupling in the Jahn – Teller inactive compound. For this purpose we choose the $HoFeO_3$ which is a Jahn – Teller inactive and may become Jahn – Teller active by substituting Cr. We anticipate that electron transfer mechanism can takes place from the Fe site to the Cr site when the incident photon energy equals to the ground state energy gap between $Fe^{3+}$ and $Cr^{3+}$ [15].

Interestingly, $HoFeO_3$ has been believed to possess the canted G-type antiferromagnetism and a potential candidate for the ultrafast recording[16] with a magnetic ordering temperature around 641 K[17]. On the other hand, the rare earth ortho-chromites of the formula $RCrO_3$ (R = Ho, Er, Yb, Lu, Y) have been believed to exhibit the multiferroicity with a canted antiferromagnetic behaviour in the temperature range of 113-140 K ($T_N$) and a dielectric transition in the temperature range of 472 - 516 K[18]. In contrary, ortho-ferrites and ortho-chromites have been believed to possess a p – type semiconducting behaviour[19], which may be very much useful in developing the optoelectronic devices. However, to the best of our knowledge, detailed information on the band gap values is not available in the literature neither on the $HoFeO_3$ nor on the $HoCrO_3$. In addition, it would also be of great interest to monitor the change in the band gap value by applying the chemical pressure either at the Fe site or the Cr site. To achieve our goal, in the present work, Cr is chosen to apply the chemical pressure at the Fe-site in $HoFeO_3$. Essentially, the fact that both the Fe and Cr are having same ionic state (in $3^+$ valence state) and the smaller ionic radius of the $Cr^{3+}$ compared with the $Fe^{3+}$ may influence the optical properties by occupying the octahedral $Fe^{3+}$ site.

As x – rays and the Raman scattering are very much sensitive for the structural changes and the local environment in an unit cell respectively, one can get precise information about the atomic positions and information about the Raman active modes. Hence, the first aim of our manuscript is to study the interplay between the electron and lattice dynamics using the Raman scattering in the HoFe$_{1-x}$Cr$_x$O$_3$ [x = 0, 0.25, 0.5, 0.75 and 1]. Secondly, we also would like to understand how the band gap value varies by applying the chemical pressure at Fe site.

**Experimental Details**:

Polycrystalline compounds of HoFe$_{1-x}$Cr$_x$O$_3$ (0 ≤ x ≤ 1) were prepared by the conventional solid state reaction method. High purity oxide powders of Ho$_2$O$_3$, Fe$_2$O$_3$, and Cr$_2$O$_3$ (purity > 99.9%) (Sigma-Aldrich chemicals India) were used as starting raw materials and were mixed together in stoichiometric ratios. The mixture thus obtained was thoroughly and repeatedly ground in the isopropanol alcohol using an agate mortar and pestle to ensure the homogeneity. Pellets were prepared using the resultant powder and sintered sequentially at 1000$^{\circ}$C for 12 h, 1200$^{\circ}$C for 12 h and 1250$^{\circ}$C for 24 h respectively. The phase purity or the structural analysis was carried out at room temperature using the powder x - ray diffraction (XRD) (Panalytical X-ray diffractometer ) with Cu K$_{\alpha}$ radiation (λ = 1.5406 Å) and with a step size of 0.017$^0$ in the wide range of the Bragg angles 2θ (20$^0$ - 80$^0$). Raman spectra was measured at room temperature using a Laser Micro Raman spectrometer (Bruker, Senterra) with an excitation source of 535 nm and with a power of 10 mW. Optical absorbance of the samples were measured at room temperature using the Perkin Elmer Lambda 1050 UV-Vis-NIR spectrophotometer in the wavelength range of 200 – 800 nm.

**Results and Discussion**:

Phase purity of the HoFe$_{1-x}$Cr$_x$O$_3$ (0 ≤ x ≤ 1) compounds is confirmed with the room temperature XRD. Intense reflections that are present in Fig. 1 are allowed reflections for a GdFeO$_3$ type disordered pervoskite structure described by the orthorhombic with a space group of the Pbnm. We do not see any impurity phase apart from the parent phase within the detectable limits of the XRD. In order to get more insights about the structural aspects, we also have performed the Rietveld refinement using the General Structural Analysis System (GSAS)[20]. Information extracted from the refinement is depicted in the Table. 1. From the Rietveld refinement data, small $\chi^2$ values of all the compounds infer that there exists a good agreement between the observed and the calculated diffraction patterns. As it is evident from the Fig. 2(a) that extracted lattice parameter values from the refinement indicates a tendency towards decrement, which is consistent with the fact that the ionic radius of the $Fe^{3+}$ (0.645 Å) is larger than that of the $Cr^{3+}$ (0.615 Å)[21]. It is worth noting that the change in the position and the shape of the diffraction peaks with Cr concentration is minimal, hinting that there exists no structural transformation as a result of the Cr dopants. This mismatch in ionic radii can leads to a distortion of the lattice and such distortion can be quantified using the Goldschmidt's tolerance factor (GTF). In general, the GTF can be defined as

$$t = \frac{r_A + r_O}{\sqrt{2}(r_B + r_O)} \qquad (1)$$

Where $r_A$, $r_B$, and $r_O$ are the radii of $Ho^{3+}$, $Fe^{3+}/Cr^{3+}$, and $O^{2-}$ respectively, where $r_B = (1-x)r_{Fe} + xr_{Cr}$. Calculated GTF for all the compounds in the present investigation are found to be in the range of 0.848 – 0.868, the range for a compound to be in an orthorhombic structure. The variation of the GTF with respect to the Cr composition is depicted in Fig. 2(b). Upon closer observation, the GTF is found to increase as a function of the chromium content, hinting that the increase in stability and tendency towards the cubic structure upon Cr doping. We have also calculated the average tilt angle $<\varphi>$ of $FeO_6$ octahedral around the pseudo cubic [111] direction using the geometric relation that has been proposed by O'Keefe and Hyde[22] and the two super exchange angles $\theta_1$ = Fe (Cr)-$O_1$-Fe (Cr), $\theta_2$ = Fe (Cr)-$O_2$-Fe (Cr). Both the above angles Fe (Cr)-$O_1$-Fe (Cr) and Fe (Cr)-$O_2$-Fe (Cr) are extracted from the structural refinement. From Fig. 2(b) it is evident that average tilt angle $<\varphi>$ of the $FeO_6$ octahedral decreases with an increase in Cr content. It is apparent that the tolerance factor increases whereas the tilt angle diminishes, hinting that the internal stresses as a result of the chemical pressure (via $Cr^{3+}$ doping) leads to a distortion of the lattice.

**Table 1:** Lattice parameters, cell volumes, selected bond lengths, bond angles from Rietveld refinement of $HoFe_{1-x}Cr_xO_3$.

| Compounds | $HoFeO_3$ | $HoFe_{0.75}Cr_{0.25}O_3$ | $HoFe_{0.5}Cr_{0.5}O_3$ | $HoFe_{0.25}Cr_{0.75}O_3$ | $HoCrO_3$ |
|---|---|---|---|---|---|
| Space group | Pbnm | Pbnm | Pbnm | Pbnm | Pbnm |
| **Lattice Parameters** | | | | | |
| a (A°) | 5.28339(6) | 5.27487(6) | 5.26645(6) | 5.25658(7) | 5.25009(8) |
| b (A°) | 5.59100(6) | 5.57407(6) | 5.55568(6) | 5.53709(7) | 5.51635(8) |
| c (A°) | 7.60986(8) | 7.59292(8) | 7.57590(8) | 7.55790(10) | 7.54373(11) |
| Cell Volume (A°³) | 224.7909 | 223.2507 | 221.6610 | 219.9814 | 218.4764 |
| **Selected bong angles (°)** | | | | | |
| Fe(Cr) – O1 – Fe(Cr) | 144.2(5) | 144.0(5) | 145.3(4) | 145.8(4) | 146.2(4) |
| Fe(Cr) – O2 – Fe(Cr) | 144.7(4) | 145.5(4) | 144.45(33) | 145.47(31) | 146.13(34) |
| **Selected bond lengths (A°)** | | | | | |
| Fe(Cr) – O1 | 1.9990(30) | 1.9959(28) | 1.9840(23) | 1.9767(21) | 1.9708(23) |
| Fe(Cr) – O2 | 2.037(9) | 2.033(8) | 2.035(8) | 2.015(6) | 2.005(7) |

Now we discuss the results pertinent to the Raman scattering which essentially gives the local structure, shift and distortion of the modes as a result of the chemical doping (here $Cr^{3+}$ doping). In addition, we also would like to correlate our structural information with the Raman data that we obtained. For this purpose the room temperature Raman spectroscopy was used on the $HoFe_{1-x}Cr_xO_3$ ($0 \leq x \leq 1$) compounds to understand the aforesaid properties.

It has been observed that in an ideal perovskite ($ABO_3$), the B-site transition metal cation locates at the centre of the oxygen octahedral and A-site cation locates at the corners of the cube[23]. However, due to the displacement of the crystallographic sites from the ideal cubic positions, most perovskites show the symmetry breaking which results the appearance of the Raman active modes in the Raman spectra. The rare earth ortho-ferrite $HoFeO_3$ is an orthorhombically distorted perovskite with a space group of $D_{2h}^{16}$ (Pbnm). The irreducible representations corresponding to the phonon modes at the Brillouin zone center[24] can be defined as follows

$$7A_g + 7B_{1g} + 5B_{2g} + 5B_{3g} + 8A_u + 8B_{1u} + 10B_{2u} + 10B_{3u}$$

Here, $A_g$, $B_{1g}$, $B_{2g}$, $B_{3g}$ are the Raman active mode species, $B_{1u}$, $B_{2u}$, $B_{3u}$ are the infrared mode species and $A_u$ is the inactive mode. Among them the modes which are above 300 cm$^{-1}$ are related to the vibrations of oxygen and the modes below the wave number 300 cm$^{-1}$ are associated with the rare earth ions[25]. However, the Raman vibrational modes corresponding to an orthorhombic structure are: $A_g + B_{1g}$ and $2B_{2g} + 2B_{3g}$, which are symmetric and antisymmetric modes respectively; In contrast, $A_g + 2B_{1g} + B_{3g}$, $2A_g + 2B_{2g} + B_{1g} + B_{3g}$, and $3A_g + B_{2g} + 3B_{1g} + B_{2g}$ are associated with the bending modes, rotation and tilt mode of the octahedral and for the changes in the rare earth movements respectively[26]. Fig. 3 shows results pertinent to the Raman Spectra of $HoFe_{1-x}Cr_xO_3$ ($0 \leq x \leq 1$) samples recorded at room temperature and with the wavenumber in the range of 50 to 800 cm$^{-1}$. Peaks with the high intensities are evident at 109, 137, 158, 337, 423 and 494 cm$^{-1}$ which is normal for a typical orthoferrite[24]. Apart from aforesaid modes, a mode at 670 cm$^{-1}$ prevailed in the compounds with the $Cr^{3+}$ ion. However, the peak which is evident at 670 cm$^{-1}$ consists very less intensity in the parent $HoFeO_3$ and $HoCrO_3$. Basically, such peak picks intensity only if we have the combination of both the Fe and Cr at the B site. Such intriguing phenomenon may be correlated to the orbital mediated electron – phonon coupling like in case of $LaFe_{1-x}Cr_xO_3$[11].

The Raman active modes of the samples are designated according to the method proposed by Gupta[27] *et. al,*. The appearance of an $A_g$ like mode with observable intensity at high frequency of around 670 cm$^{-1}$ can be attributed to an in-phase stretching (breathing) mode of oxygen in the close vicinity of the substituted $Cr^{3+}$ ion. Essentially the oxygen breathing mode is activated by the charge transfer between the $Fe^{3+}$ and $Cr^{3+}$ through an orbital mediated electron-phonon coupling mechanism[11].

In $HoFe_{1-x}Cr_xO_3$ compounds, the ground state electron configuration of $Fe^{3+}$ does not support orbital mediated electron-phonon coupling due to the lack of strongly interacting half-filled $e_g$ levels in both the $Fe^{3+}$ ($d^5$) and $Cr^{3+}$ ($d^3$). To facilitate such an orbital mediated electron-phonon coupling, the electronic configuration of $Fe^{3+}$ must contain partially filled $e_g$ orbital (like $Fe^{4+}$) such that $d^4$ electron of the $Fe^{4+}$ ion can move to the upper $e_g$ levels of the $Cr^{2+}$ to create electronic excitation as shown in $LaMnO_3$[28]. Fig. 4 explains the mechanism for the orbital mediated electron-phonon coupling. The left part of Fig. 4(a) shows electronic states of $Fe^{3+}$ and $Cr^{3+}$ ions. $v_{g,0}$, $v_{g,1}$, $v_{g,2}$, …., $v_{g,n}$ and $v_{e,0}$, $v_{e,1}$, $v_{e,2}$, …, $v_{e,n}$ represents the vibrational states of the $Fe^{3+}$ and the $Cr^{3+}$ respectively. Right part of the Fig. 4(a) reveals electronic states of the $Fe^{4+}$ and $Cr^{2+}$ ions. Arrows on both Fig. 4(a) and 4(b) represents various transitions between $Fe^{3+} \rightarrow Cr^{3+}$ and $Fe^{4+} \rightarrow Cr^{2+}$ respectively.

It has been reported that the photon mediated charge transfer can takes place between the $Fe^{3+}$ and $Cr^{3+}$ ions upon irradiation with a laser of wavelength 535 nm[11, 15]. In this process, the overlap between the *d*-orbitals of $Cr^{2+}$ and the *p*-orbitals of the oxygen couples through a lattice distortion, causing a self-trapping motion. Evidently, this motion increases the lifetime of excited $Cr^{2+}$ electronic ground state long enough to interact with the intrinsic phonon mode. Essentially, during the charge transfer mechanism (CT), when the photon energy equals to the CT energy gap between the two transition metal ions $Fe^{3+}$ and $Cr^{3+}$, electrons in the $Fe^{3+}$ excite to the $Cr^{3+}$ ion and leaves them in a strongly coupled $d^4$-$d^4$ configuration with the half-filled bands. The change in the charge density of $e_g$ orbital of transition metal surrounded by the oxygen octahedral activates a breathing distortion of $O_6$ around the transition metal cation which appears in the Raman spectrum at around 670 cm$^{-1}$. This configuration is Jahn-Teller active, hence, that it leads to a volume preserving lattice distortion (δ), involves a stretching of Fe (Cr)-O bonds along *z* - direction and a compression in x-y plane. As a result of the Jahn - Teller effect, an electron in $e_g$ orbital collapse into the lower energy state which produces a potential minimum. This minimum potential traps the electron in that orbital (self-trapping) and increases its life time in the excited $Cr^{2+}$ state long enough for it to interact with intrinsic phonon mode/lattice distortion. In the perturbed state, indeed there exists a contraction of the oxygen octahedral surrounding to the $Fe^{4+}$ ion and leads to an expansion in the adjacent octahedron surrounding to $Cr^{2+}$ ion as shown in Fig. 4(b). The oxygen lattice relaxes back to its unperturbed state when the electron transfers back to $Fe^{3+}$ state. In this fashion, the charge transfer of an electron from $Fe^{3+}$ to the $Cr^{3+}$ ion activates an oxygen breathing mode of $A_g$ symmetry through orbital mediated electron-lattice coupling.

The increase in the intensity of the peak at around 670 cm$^{-1}$ is observed only in doped samples as shown in Fig. 5 (a). This effect can be related to the increase in the degree of disorder, which can be supported by an increase in the tolerance factor as a function of the Cr content. This is a striking coincidence between our structural and Raman studies. As the Cr content increases, there would indeed be an increase in the degree of disorder, which may enhance the interaction between the lattice distortion and the charges transferred between $Fe^{3+}$ - $Cr^{3+}$. Eventually as a result of this there would be an enhancement in the electron-phonon coupling which leads to increase in the intensity of the peak at 670 cm$^{-1}$. The broadening of the peak at 670 cm$^{-1}$ with an increase in the amount of the $Cr^{3+}$ ions at the $Fe^{3+}$ site can correlate with the structural disorder. This is confirmed by the observed change in the lattice parameters by the substitution of the $Cr^{3+}$ ions at the Fe site. The observed shift in the wave number towards the higher values (Blue shift) in the doped compounds shown in Fig 5(b) can correlate to the compressive strain produced in the compounds by the incorporation of the $Cr^{3+}$ ion at the Fe site. This effect is supported by the change in Fe (Cr) – O bond lengths as well as $FeO_6$ octahedral tilt angle with respect to the chromium content at Fe site as shown in the Table **1**.

The optical absorbance of the $HoFe_{1-x}Cr_xO_3$ (0 ≤ x ≤ 1) compounds was recorded at room temperature and is shown in Fig. 6. Indeed there exists a direct band gap and the value of gap is determined using the Tauc's equation[29]. Essentially this equation relates the optical absorption coefficient (α), photon energy ($h\nu$) and the energy gap $E_g$ as given below

$$\alpha h\nu = \left(h\nu - E_g\right)^{1/2}$$

Optical band gaps of the $HoFe_{1-x}Cr_xO_3$ (0 ≤ x ≤ 1) compounds are obtained using the above equation and by extrapolating the linear region of the curve to the zero in the $(\alpha h\nu)^2$ vs. $h\nu$ graph as shown in Fig. 7 (a). The direct band gap value for $HoFeO_3$ and $HoCrO_3$ is calculated as 2.07 eV and 3.26 eV respectively. From the Fig. 7(b) and for the compounds with the combination of Fe and Cr, it is evident that the band gap decreases with Cr content, and reached a minimum value of 1.94 eV at $x = 0.5$. Further increase in Cr content results in increase of the band gap value and reached a maximum value of 3.26 eV at $x = 1$. In order to explain the observed band gap variation we would like to propose a possible mechanism using energy diagram of $HoFeO_3$, $HoCrO_3$ and $HoFe_{1-x}Cr_xO_3$. Fig. 8 (a) shows the energy diagram of $HoFeO_3$ with the band gap value of 2.07 eV (b) the energy diagram of $HoCrO_3$ with the bad gap value of 3.26 eV and (c) probable energy diagram for $HoFe_{1-x}Cr_xO_3$ for which the band gap varies between 2.07 – 3. 26 eV. The variation of the band gap with Cr concentration could be due to a complex interplay between the $Fe^{3+}$ and the $Cr^{3+}$ electronic levels mediated by oxygen through superexchange interaction. From the Fig. 8(c) it is evident that when x < 0.5, the valance band maxima (VBM) and conduction band minima (CBM) shifts to higher energy (dark red colour). The shift in VBM may be explained on the basis of the hybridization of $d$ - orbitals of Fe and Cr with $p$ - orbitals of oxygen in the valance band. Essentially, the ferrimagnetically[15] coupled $Fe^{3+}$ and $Cr^{3+}$ induce a spin disorder on oxygen which can enhance the broadening of oxygen $p$ - orbitals and valance band edges of $Fe^{3+}$ and $Cr^{3+}$ [30, 31], hinting a smaller band gap until x = 0.5. When x > 0.5, the width of available un occupied $d$ - orbitals of $Fe^{3+}$ at the conduction band reduces, which can leads to a shift of the conduction band minima to higher energies (as shown in fig 8(c) (purple colour)). As a result, the band gap in $HoFe_{1-x}Cr_xO_3$ increases above x = 0.5 and reaches a maximum value of 3.26 eV at x = 1. From the above results, indeed it is possible to tune the band gap in rare earth ortho-ferrites and the other compounds with a similar structure by controlling the Fe/Cr ratio.

In summary, we have explored the orbital mediated electron – phonon coupling mechanism in the compounds $HoFe_{1-x}Cr_xO_3$ (0 ≤ x ≤ 1). There is a striking coincidence between our structural and Raman studies. Raman studies infer that, $A_g$ like symmetric oxygen breathing mode at around 670 cm$^{-1}$ in the compounds with both the Fe and Cr. The decrease in optical band gap is ascribed to the induced spin disorder due to the $Fe^{3+}$ and the $Cr^{3+}$ on oxygen, which can leads to broadening of the oxygen $p$-orbitals. On the other hand, increase in band gap value explained on the basis of the reduction in the available unoccupied $d$-orbitals of $Fe^{3+}$ at the conduction band. The present results would indeed be helpful in understanding and to develop optoelectronic devices based on orthoferrites.

We would like to acknowledge Indian Institute of Technology, Hyderabad and Department of Science and Technology (DST) **(**Project #SR/FTP/PS-190/2012**)** for the financial support.

**References**:

(1) M. Eibschütz, S. Shtrikman, and D. Treves, *Phys. Rev.* **156**, 562 (1967).

(2) R. L. White, *J. Appl. Phys.* **40**, 1061 (1969).

(3) N. S. Ovanesyan and V. A. Trukhtanov, *JETP Lett.* Engl. Transl. **17**, 67 (1973).

(4) M. Gateshki, L. Suescun, S. Kolesnik, J. Mais, K. Świerczek, S. Short, B. Dabrowski, *Journal of Solid State Chemistry* **181**, 1833 (2008).

(5) N. Singh, J. Y. Rhee, S. Auluck, *J. Korean Phys. Soc.* **53**, 806 (2008).

(6) Xiutao Ge, Yafei Liu, Xingqin Liu, *Sensors and Actuators B* **79**, 171 (2001).

(7) Yuuki Hosoya, Yoshiteru Itagaki, Hiromichi Aono, Yoshihiko Sadaoka, *Sensors and Actuators B* **108**, 198 (2005).

(8) Xinshu Niu, Honghua Li, Guoguang Liu, *Journal of Molecular Catalysis A: Chemical* **232**, 89 (2005).

(9) S. Acharya, J. Mondal, S. Ghosh, S.K. Roy, P.K. Chakrabarti, *Materials Letters* **64**, 415 (2010).

(10) A. P. Ramirez, *J. Phys.: Condens. Matter* **9**, 8171 (1997).

(11) Jacob Anderasson, Joakim Holmund, Christopher S. Knee, Mikael Käll, Lars Börjesson, Stefan Naler, Joakim Bäckström, Michael Rübhausen, Abul Kalam Azad, and Sten-G. Eriksson, *Phys. Rev. B* **75**, 104302 (2007).

(12) V. Gnezdilov, P. Lemmens, Yu. G. Pashkevich, Ch. Payen, K. Y. Choi, J. Hemberger, A. Loidl, and V. Tsurkan, *Phys. Rev. B* **84**, 045106 (2011).

(13) P. B. Allen and V. Perebeinos, *Phys. Rev. Lett.* **83**, 4828 (1999).

(14) A. Dubroka, J. Humliček, M. V. Abrashev, Z. V. Popović, F. Sapiña, and A. Cantarero, *Phys. Rev. B* **73**, 224401 (2006).

(15) K. Miura and K. Terakura, *Phys. Rev. B* **63**, 104402 (2001).

(16) A. V. Kimel, B. A. Ivanov, R. V. Pisarev, P. A. Usachev, A. Kirilyuk and Th. Rasing, *Nature Physics* **5**, 727 (2009).

(17) Zhiqiang Zhou, Li Guo, Haixia Yang, Qiang Liu, Feng Ye, *Journal of Alloys and Compounds* **583**, 21 (2014).

(18) Jyoti Ranjan Sahu, Claudy Rayan Serrao, Nirat Ray, Umesh V. Waghmare and C. N. R. Rao, *J. Mater. Chem.* **17**, 42 (2007).

(19) Maike Siemons, Ulrich Simon, *Sensors and Actuators B* **126**, 181 (2007).

(20) A.C. Larson and R.B. Von Dreele, *General Structure Analysis System* (GSAS), Los Alamos National Laboratory Report LAUR 86–748 (2004).

(21) R.D.Shannon, *Acta Cryst. A* **32**, 751 (1976).

(22) Tapan Chatterji, Bachir Ouladdiaf, P. Mandal, B. Bandyopadhyay, and B. Ghosh, *Phys. Rev. B* **66**, 054403 (2002).

(23) M. A. Peña and J. L. G. Fierro, *Chem Rev.* **101**, 1981 (2001).

(24) S. Venugopalan, M. Dutta, A. K. Ramdas, J. P. Remeika, *Phys. Rev. B* **31**, 1490 (1985).

(25) Manoj K. Singh, Hyun M. Jang, H. C. Gupta and Ram S. Katiyar, *J. Raman Spectrosc.* **39**, 842 (2008).

(26) M. N. Iliev, M. V. Abrashev, H. -G. Lee, V. N. Popov, Y. Y. Sun, C. Thomsen, R. L. Meng, C. W. Chu, *Phys. Rev. B* **57**, 2872 (1998).

(27) H. C. Gupta, Manoj Kumar Singh and L. M. Tiwari, *J. Raman Spectrosc.* **33**, 67 (2002).

(28) Vasili Perebeinos and Philip B. Allen, *Phys. Rev. B* **64**, 085118 (2001).

(29) Anders hagfeldt and Michael Grätzel, *Chem. Rev.* **95**, 49 (1995).

(30) T. Arima, Y. Tokura, and J.B. Torrance, *Phys. Rev. B* **48**, 17006 (1993).

(31) Yong Wang, Kenneth Lopata, Scott A. Chambers, Niranjan Govind, and Peter V. Sushko, *J. Phys. Chem. C* **117**, 25504 (2013).

**Figure captions**

**Fig. 1:** Powder x-ray diffraction patterns of HoFe$_{1-x}$Cr$_x$O$_3$ ($0 \leq x \leq 1$). It is evident that all the compounds are formed in single phase.

**Fig. 2:** (a) Variation of lattice parameter with Cr composition. (b) Variation of tolerance factor (circle symbol) and FeO$_6$ average tilt angle (square symbol) with Cr composition.

**Fig. 3:** Room temperature Raman spectra of HoFe$_{1-x}$Cr$_x$O$_3$ ($0 \leq x \leq 1$) compounds with an excitation of 535 nm.

**Fig. 4:** (a) Franck-Condon (FC) mechanism for Jahn-Teller active perovskites. $\nu_{g,0}$, $\nu_{g,1}$, $\nu_{g,2}$, …., $\nu_{g,n}$ and $\nu_{e,0}$, $\nu_{e,1}$, $\nu_{e,2}$, …, $\nu_{e,n}$ represents the vibrational states of Fe$^{3+}$ and Cr$^{3+}$ respectively. For FC mechanism to happen for a vibrational mode, the virtual state $|\nu_r\rangle$ of Raman process must coincide with any vibrational state of electronically excited state. δ indicates lattice distortion due to Jahn - Teller effect as a result of charge transfer mechanism (b) Octahedral sites of Fe$^{4+}$ and Cr$^{2+}$ respectively. Dotted arrow in the figure indicates charge transfer mechanism and lattice relaxation.

**Fig. 5:** (a) Intensity variation of A$_g$ peak and (b) wave number shift of A$_g$ peak for the HoFe$_{1-x}$Cr$_x$O$_3$ ($0 \leq x \leq 1$) compounds.

**Fig. 6:** Absorption spectra of HoFe$_{1-x}$Cr$_x$O$_3$ ($0 \leq x \leq 1$) compounds.

**Fig. 7:** (a) Tauc's plots to determine the band gap values of HoFe$_{1-x}$Cr$_x$O$_3$ ($0 \leq x \leq 1$) compounds. (b) Variation of the band gap with respect to Cr composition.

**Fig. 8:** (a) Shows the energy diagram of HoFeO$_3$ (b) Energy diagram of HoCrO$_3$ (c) probable energy diagram for HoFe$_{1-x}$Cr$_x$O$_3$. It is evident from the frame (c) that when $x < 0.5$, the valance band maxima (VBM) and conduction band minima (CBM) shifts to higher energy (dark red color). However, the shift in VBM is due to strong hybridization of d orbitals of Fe & Cr with *p* - orbitals of oxygen in valance band. When $x > 0.5$, band gap is dominated by unoccupied *d* - orbitals of Cr in conduction band which leads to increase in band gap (purple color).

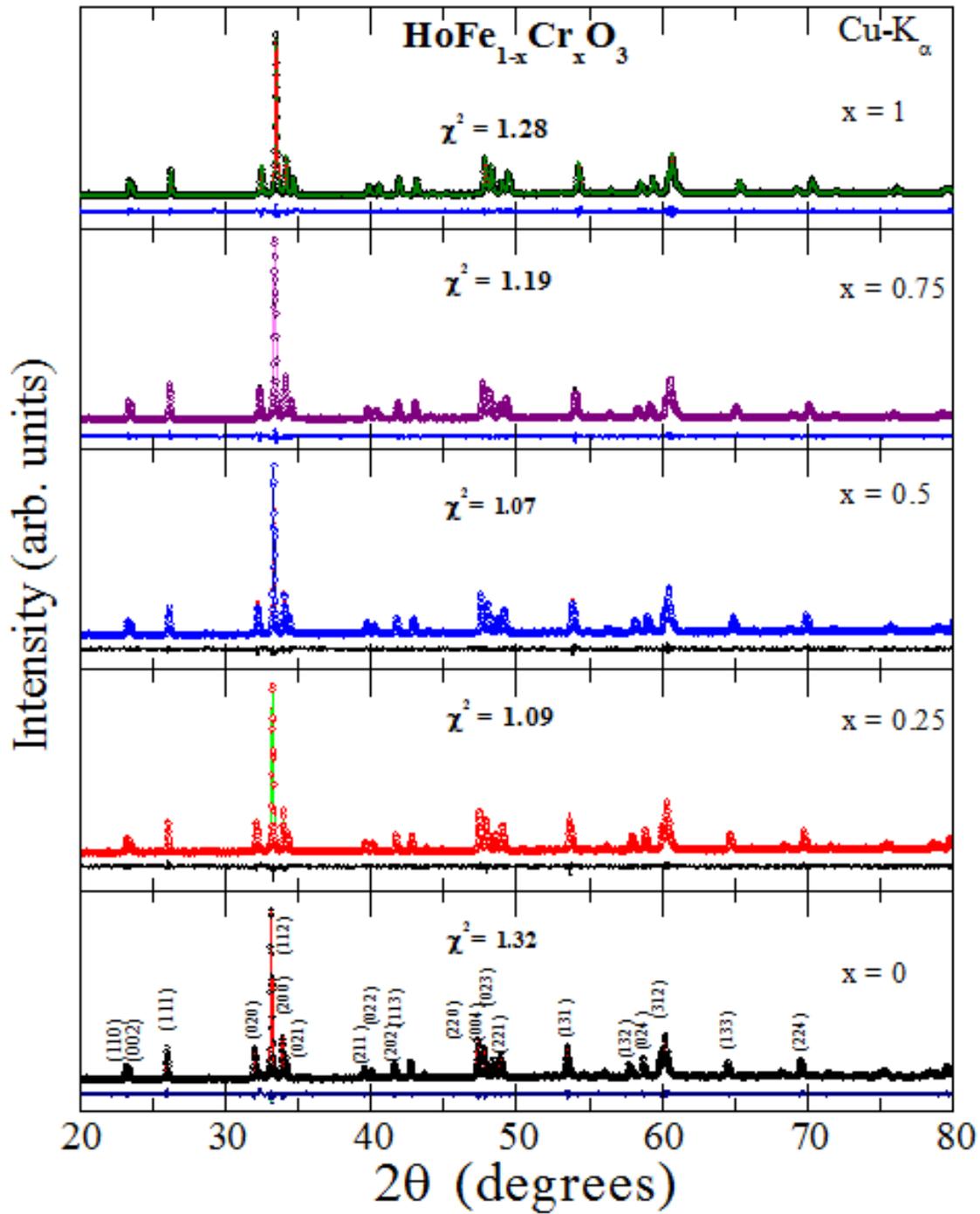

**Fig. 1:** Powder x-ray diffraction patterns of HoFe$_{1-x}$Cr$_x$O$_3$ ($0 \leq x \leq 1$). It is evident that all the compounds are formed in single phase.

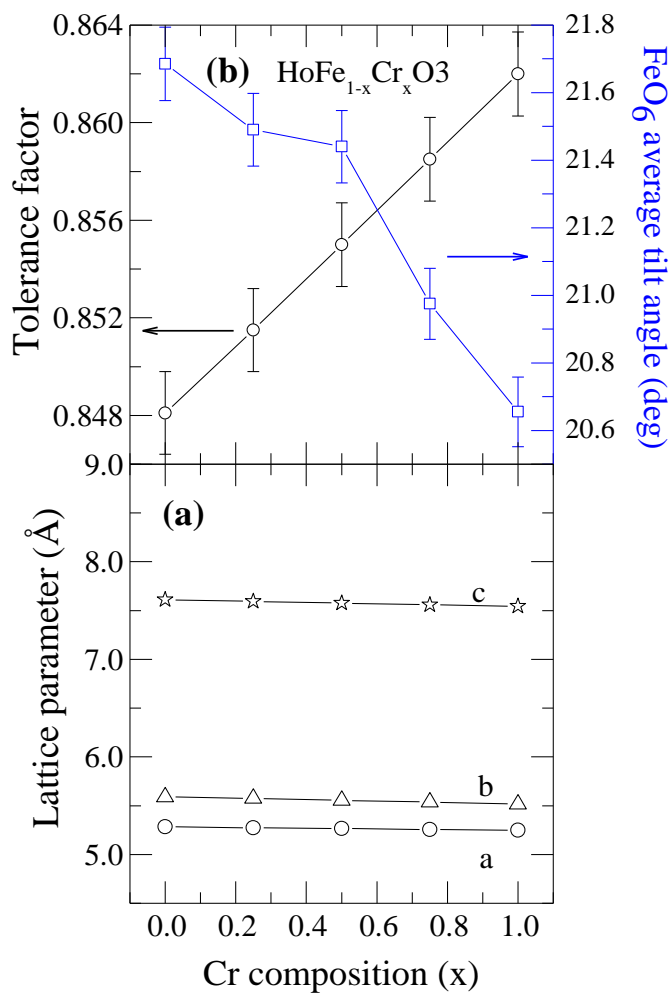

**Fig. 2:** (a) Variation of lattice parameter with Cr composition. (b) Variation of tolerance factor (circle symbol) and FeO$_6$ average tilt angle (square symbol) with Cr composition.

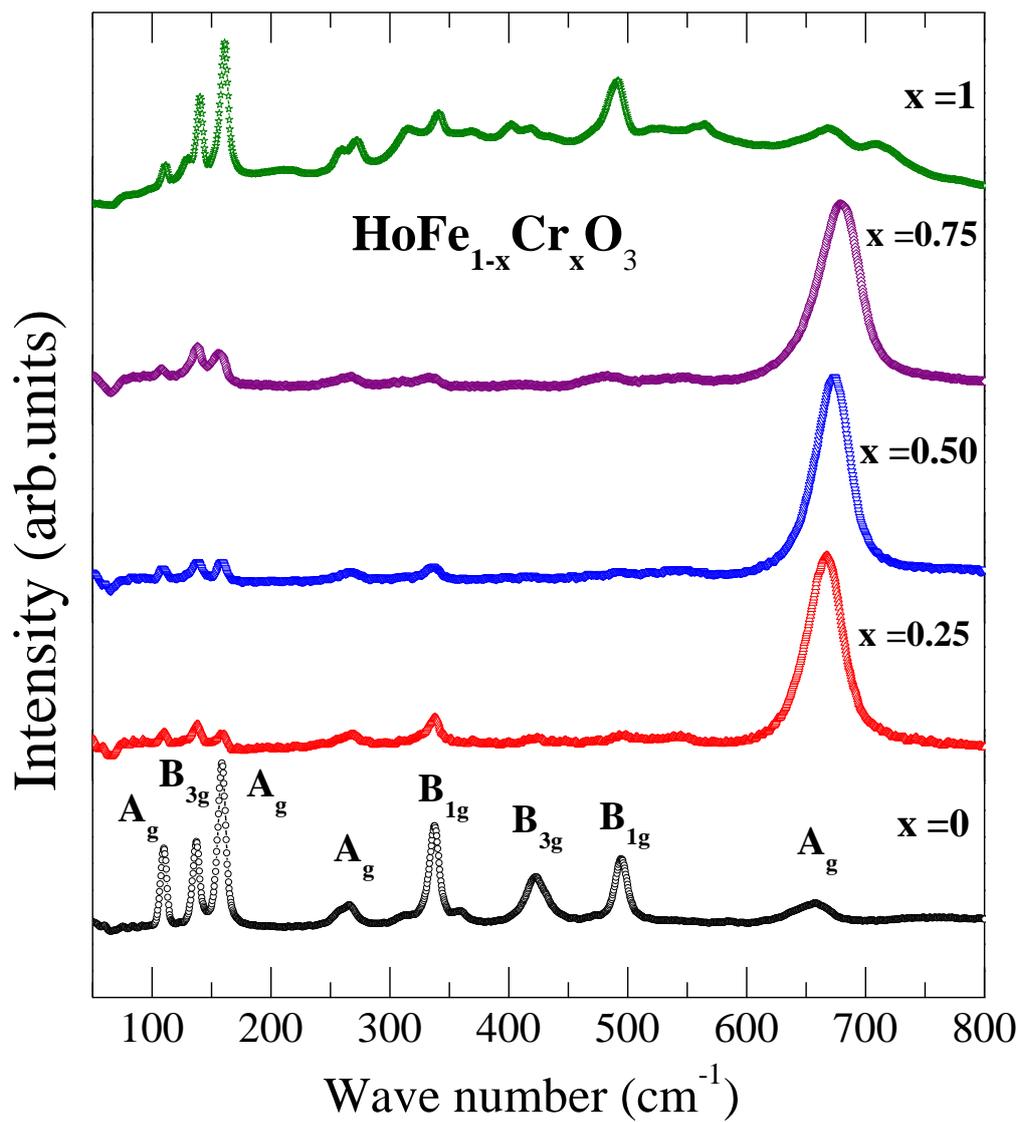

**Fig. 3:** Room temperature Raman spectra of HoFe$_{1-x}$Cr$_x$O$_3$ ($0 \leq x \leq 1$) compounds with an excitation of 535 nm.

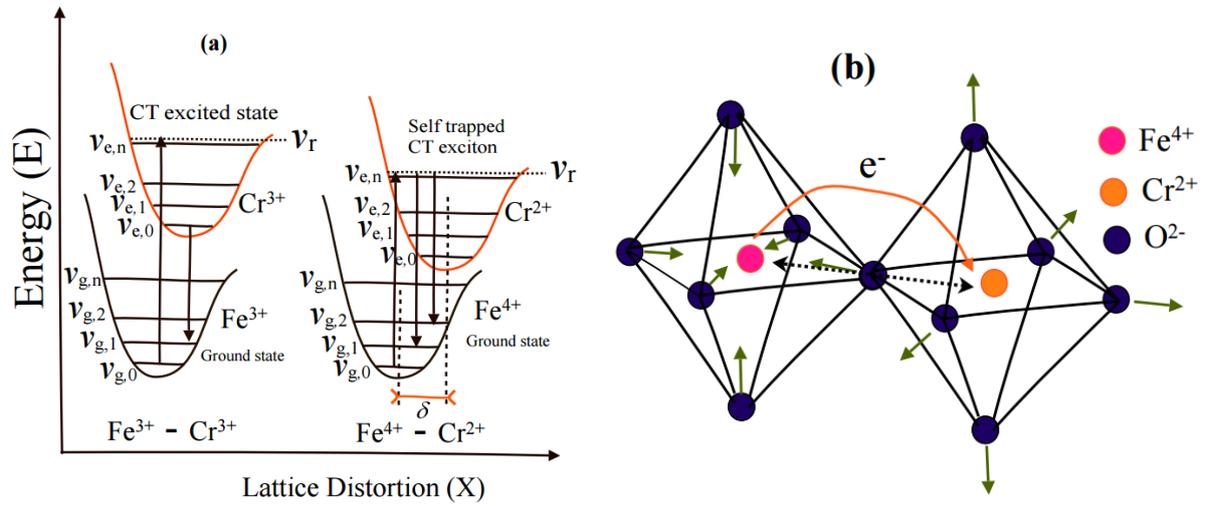

**Fig. 4:** (a) Franck-Condon (FC) mechanism for Jahn-Teller active perovskites. $v_{g,0}, v_{g,1}, v_{g,2}, ...., v_{g,n}$ and $v_{e,0}, v_{e,1}, v_{e,2}, ..., v_{e,n}$ represents the vibrational states of $Fe^{3+}$ and $Cr^{3+}$ respectively. For FC mechanism to happen for a vibrational mode, the virtual state $|v_r\rangle$ of Raman process must coincide with any vibrational state of electronically excited state. $\delta$ indicates lattice distortion due to Jahn - Teller effect as a result of charge transfer mechanism (b) Octahedral sites of $Fe^{4+}$ and $Cr^{2+}$ respectively. Dotted arrow in the figure indicates charge transfer mechanism and lattice relaxation.

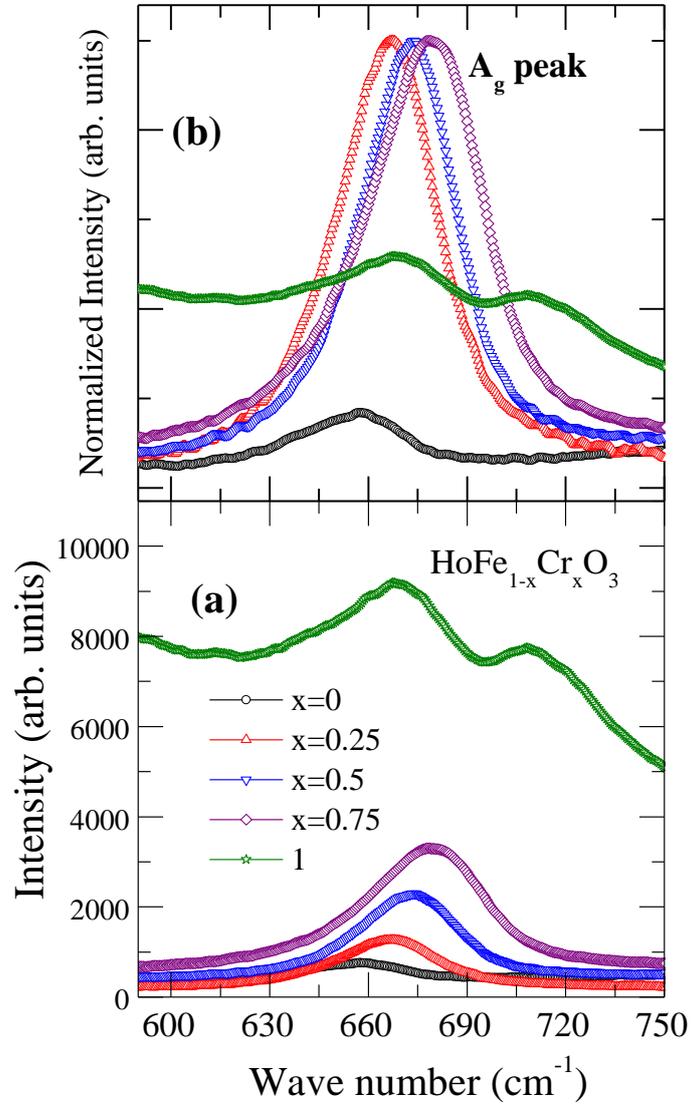

**Fig. 5:** (a) Intensity variation of $A_g$ peak and (b) wave number shift of $A_g$ peak for the HoFe$_{1-x}$Cr$_x$O$_3$ ($0 \leq x \leq 1$) compounds.

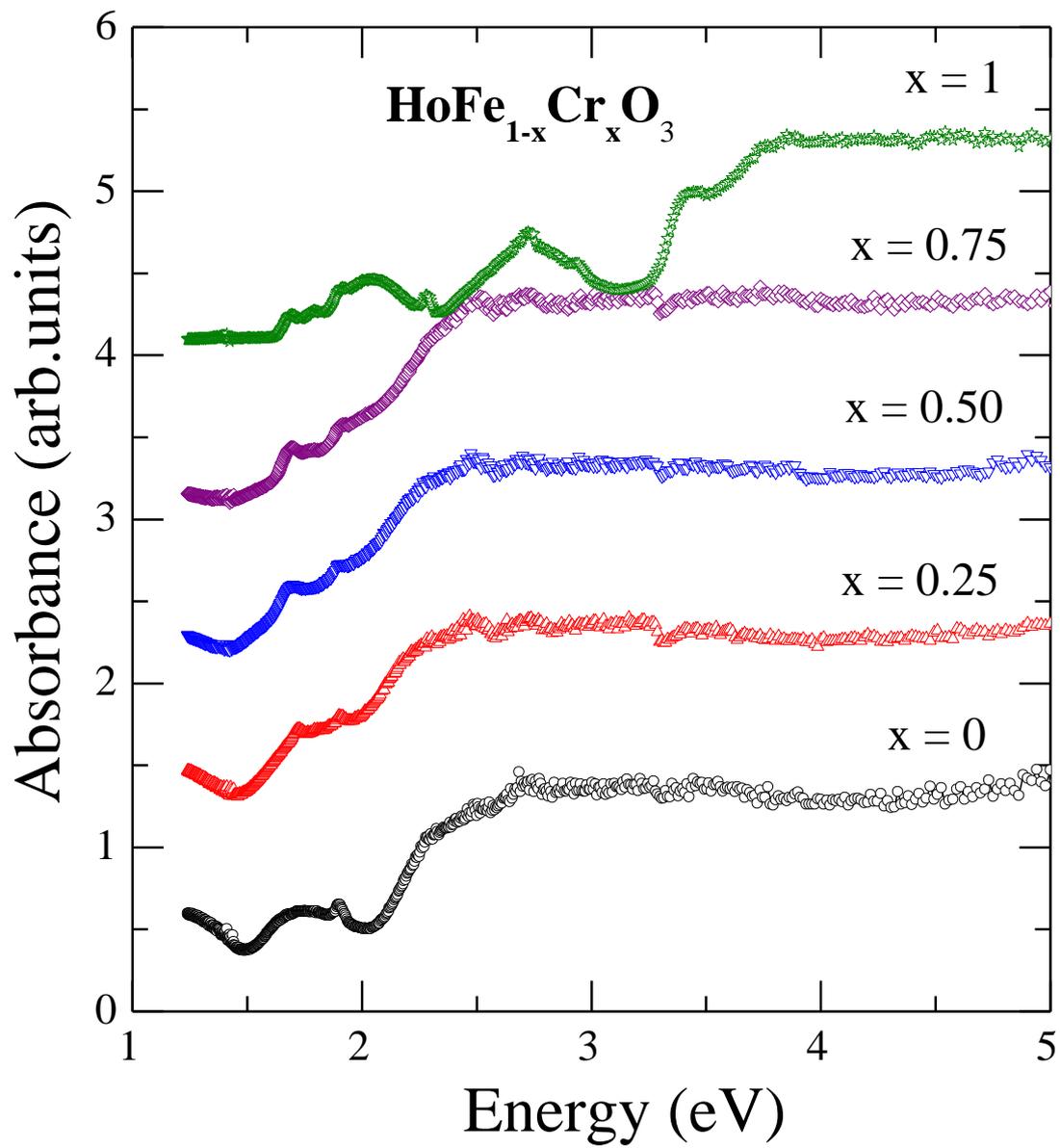

**Fig. 6:** Absorption spectra of HoFe$_{1-x}$Cr$_x$O$_3$ ($0 \leq x \leq 1$) compounds.

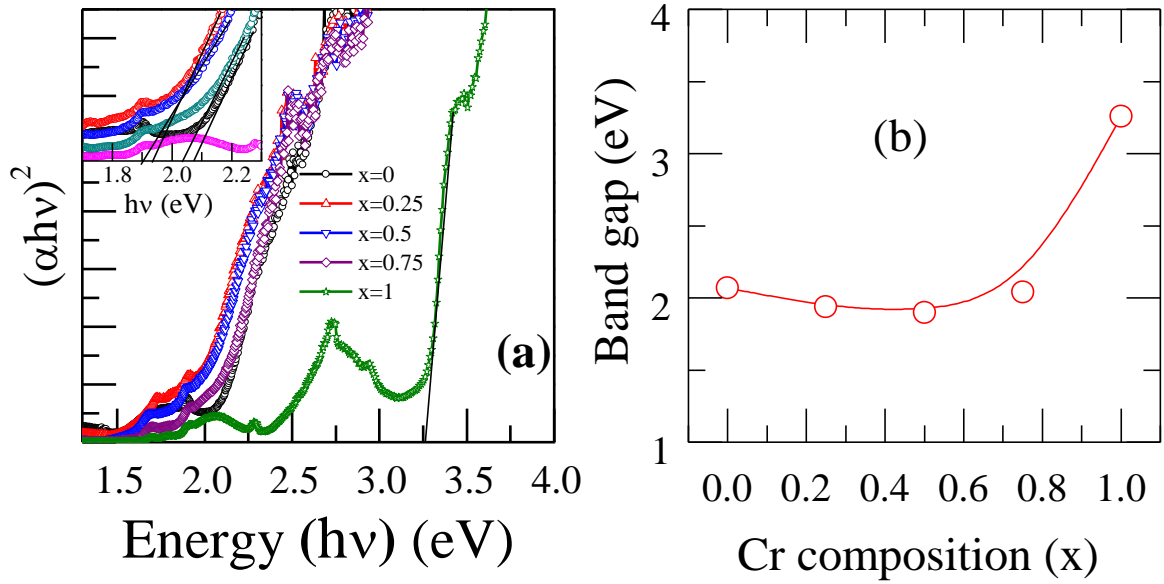

**Fig. 7:** (a) Tauc's plots to determine the band gap values of $HoFe_{1-x}Cr_xO_3$ ($0 \leq x \leq 1$) compounds. (b) Variation of the band gap with respect to Cr composition.

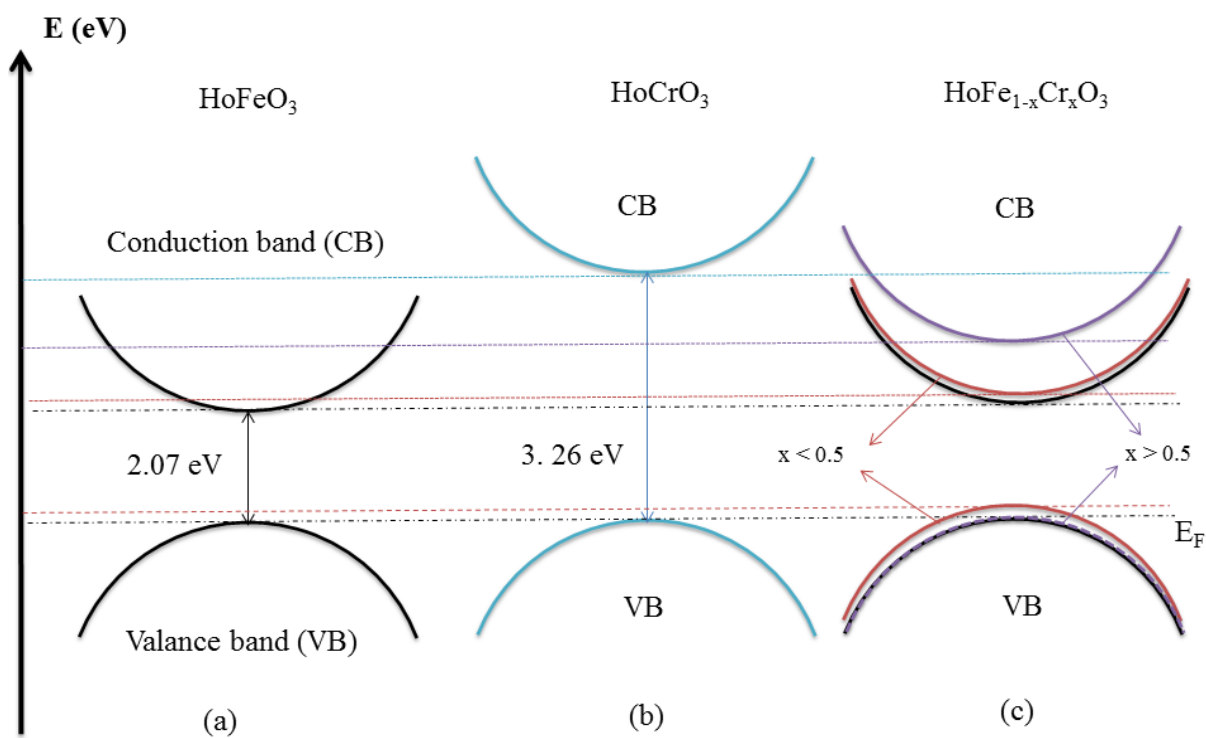

**Fig. 8:** (a) Shows the energy diagram of $HoFeO_3$ (b) Energy diagram of $HoCrO_3$ (c) probable energy diagram for $HoFe_{1-x}Cr_xO_3$. It is evident from the frame (c) that when $x < 0.5$, the valence band maxima (VBM) and conduction band minima (CBM) shifts to higher energy (dark red color). However, the shift in VBM is due to strong hybridization of d orbitals of Fe & Cr with *p* - orbitals of oxygen in valance band. When $x > 0.5$, band gap is dominated by unoccupied *d* - orbitals of Cr in conduction band which leads to increase in band gap (purple color).